\documentclass[manuscript]{aastex}

\shorttitle{Stellar flaring}
\shortauthors{Kalberla et al.}

\begin{document}

%% LaTeX will automatically break titles if they run longer than
%% one line. However, you may use \\ to force a line break if
%% you desire.

\title{Does the stellar distribution flare? \\
  A comparison of stellar scale heights with LAB {H\,{\sc i}} data}

%% Use \author, \affil, and the \and command to format
%% author and affiliation information.
%% Note that \email has replaced the old \authoremail command
%% from AASTeX v4.0. You can use \email to mark an email address
%% anywhere in the paper, not just in the front matter.
%% As in the title, use \\ to force line breaks.

\author{P.\,M.\,W. Kalberla, J. Kerp and L. Dedes}
\affil{Argelander-Institut f\"ur Astronomie, Universit\"at Bonn}

%\author{L. Dedes}
%\affil{Argelander-Institut f\"ur Astronomie, Universit\"at Bonn}
\email{pkalberla@astro.uni-bonn.de}

\and

\author{U. Haud}
\affil{Tartu Observatory, 61602 T\~oravere, Estonia}

\begin{abstract}
  The question, whether the stellar populations in the Milky Way take
  part in flaring of the scale heights as observed for the {H\,{\sc i}}
  gas is a matter of debate. Standard mass models for the Milky Way
  assume a constant scale height for each of the different stellar
  distributions. However, there is mounting evidence that at least some
  of the stellar distributions reach at large galactocentric distances
  high altitudes that are incompatible with a constant scale
  height. We discuss recent observational evidence for stellar flaring
  and compare it with H\,{\sc i} data from the Leiden/Argentine/Bonn
  (LAB) survey. Within the systemic and statistical uncertainties we
  find a good agreement between both.

\end{abstract}


\section{Introduction}
\label{into}

Observations of the {H\,{\sc i}} gas distribution in the Milky Way
galaxy have shown early that the {H\,{\sc i}} distribution is warped
\citep{Westerhout1962}. Furthermore the scale height of the density
distribution increases systematically with galactocentric radius $R$
\citep{Lozinskaya1963}. Because of the strength of this effect this
phenomenon was described as flaring.  Yet, for the stars the situation
appeared to be different. A constant scale height was reported by most
of the observers.

Flaring of {H\,{\sc i}} gas is a natural phenomenon revealing the
radial mass distribution of the Milky Way galaxy. Considering the
barometric equation, the gravitational force balances gas pressure
and turbulent motions. Pressure and turbulence can be characterized
by the velocity dispersion $\sigma_z$ of the objects under consideration
as a function of $R$. It is well established fact that the surface
density distributions of stars and gas in most of the galaxies decrease
exponentially with $R$ \citep{Freeman1970,Bigiel2012}. Consequently, the
gravitational forces $k_z(R)$ perpendicular to the plane decrease with
$R$. Since the velocity dispersion of the {H\,{\sc i}} gas distribution
is approximately independent on $R$ \citep[e.g.][]{Spitzer1968} the
associated scale height $h_z(R)$ of the gas increases exponentially with
$R$.

\cite{Lozinskaya1963} were the first to describe warp and flaring of the
{H\,{\sc i}} gas in detail. They also noted that stars do not share the
gaseous flaring and deduced from this different behavior that flaring
cannot be caused by gravitational effects. Alternatively they proposed
that regular magnetic fields cause the gaseous flaring. Today we know
that the magnetic fields are oriented predominantly parallel to the
Galactic plane, on average with a strength of a few $\mu$G
\citep{Beck2013}.  X-shaped halo fields are sometimes observed,
indicating an opening of the field lines with $R$ similar to the flaring
{H\,{\sc i}} layer. An unambiguous proof that magnetic forces cause the
{H\,{\sc i}} layer to flare is still missing \citep{Gressel2013}.

Next to the distribution of the baryonic matter flaring might be also
affected by the dark matter distribution.  \cite{Olling1995}
demonstrated that the shape of the flaring {H\,{\sc i}} layer is
affected by the flattening of the dark matter component.
\cite{OM00,OM01}, later \cite{Narayan2002,Kalberla2003,Narayan2005} and
\cite{Kalberla2007} applied this basic idea to model the Milky Way mass
distribution using full-sky {H\,{\sc i}} data. According to these
investigations it became evident that also stellar flaring is expected
to be observed \citep{Jog2007}.  \cite{OM00,OM01} assumed constant scale
heights for the stellar distribution and note that their results do not
depend on details of the vertical distribution. Contrary to that
\cite{Kalberla2003} and \cite{Kalberla2007} found interrelations between
the scale heights of all model components. Vertical changes in the mass
distribution affect the scale heights of the other components. Their aim
was eventually to match the mass model to the observed {H\,{\sc i}}
flaring. Several axisymmetric spheroidal halo models were considered but
neither oblate nor prolate models fitted at the same time all
observations; the solutions were found to depend on the radial scales
considered. The best fit was obtained with an isothermal dark matter
distribution in a thick disk and an associated ring in the Galactic
plane at $13 < R < 18.5$ kpc. Such an axisymmetric model is certainly
only an approximation because it is evident from the {H\,{\sc i}}
distribution that towards the southern sky (approximately in direction
to the Magellanic System) there is more gas and baryonic mass
\citep{Kalberla2007,Kalberla2008}. These findings agree well with the
orientation of the minor axis of the triaxial mass distribution derived
by \cite{Law2009}.

So-far the stellar distributions appear to behave differently. The
constant scale height of stars \citep{vanderKruit1982} demands an
exponential fall-off of the velocity dispersion proportional to $R$
\citep{Lewis1989}.  This, however, is hard to understand physically
since the radial decrease in surface density needs precisely to be
compensated by the fall-off of the stellar vertical velocity dispersion
$\sigma_z$ \citep[][see Sect. 3.2.3 for discussion]{vanderKruit2011}.

For most studies of stellar population synthesis flaring is either
not considered or only minor corrections of $ k_{flare} = 1 + 0.054 \,
(R - R_{flare}) $ to the stellar scale height for $ R > R_{flare} $ with
$ R_{flare} = 9.5 $ kpc are applied \citep{Gyuk99,Reyle2009}. Recently
\cite{Polido2013} used the relation $k_{exp} = e^{0.4 ( R -
  R_{\odot})/R_{\odot}}$. % where $R_\odot$ denotes the galactocentric distance of the Sun.

\section{Evidence for flaring from observations}
\label{Sectobs}

\subsection{Preliminaries}

The determination of scale heights depends on the determination of
distances to the objects under consideration. The distance scale is tied
to $R_\odot$. Here we use $R_\odot= 8 $ kpc, the standard value used
for stellar data in the publications mentioned below.

From the barometric equation, the flaring amplitude at a particular
position depends on the velocity dispersion of the component under
investigation. Accordingly, to quantify flaring, one needs to
distinguish between the underlying disk populations (either thin or
thick) since they are characterized by different velocity
dispersions. Both components may have different radial scale lengths and
a mix could therefore mimic a pseudo-flaring. We distinguish between
thin and thick disk components in the further discussion.

{H\,{\sc i}} distances are derived as kinematical distances and depend
on the rotation curve used, hence on the Galactic mass model. For
stellar distance determination photometric distances are mostly used.
Using the luminosity function
\citep{Wainscoat1992} the differential star counts along the line of
sight can be converted to space densities. Star counts need to be
corrected for extinction. The completeness of the sample and selection
effects need to be investigated carefully. Moreover one needs to
investigate potential signatures for a warp. Such details have been
taken into account by the publications discussed below but will not be
subject of our paper.

\subsection{{H\,{\sc i}} and CO}
Observations of the {H\,{\sc i}} gas in the outer part of the Milky Way
have shown early on significant changes of the scale heights with radius
$R$. \cite{Burton1976} derived for $ R > 7.6$ kpc the flaring relation
$k_{HI} = 1 + 0.19 \, (R-7.6)$. For $R = 2 R_\odot$ the flaring in
{H\,{\sc i}} is about a factor of two stronger compared to heights
used by \citet[][see Sect. \ref{intro}]{Gyuk99,Reyle2009,Polido2013}.

Also the molecular gas distribution in the Milky Way shows flaring
consistent with H\,{\sc i} data \citep[][Fig. 15]{Kalberla2007}. 
\citet{Bronfman1988,Wouterloot1990} and
\citet{Malhotra1994} analyzed CO as a tracer of the molecular gas
distribution.

In the following we will consider more recent {H\,{\sc i}} LAB survey data
\citep{Kalberla2005} and flaring compilations from
\cite{Kalberla2007,Kalberla2008,KK2009}. 

\subsection{Cepheids}
\citet[][F14]{Feast2014} report the detection of five classical Cepheid stars
towards the Galactic bulge at $ 1 - 2 $ kpc above the
plane. Distances were derived from the well-calibrated period-luminosity
relationship. These stars were observed in two colors, so that their
distances and reddenings could be determined simultaneously. The puzzling
result however is that these massive young stars (age $\leq 130\cdot
10^6$\,years) are located such high above the Galactic plane. Because
they are observed towards both Galactic hemispheres this finding is
interpreted as a strong indication for stellar
flaring at distances of $ 13 < R < 22$ kpc. 
\cite{Feast2014} conclude that the derived heights are consistent with
{H\,{\sc i}} flaring according to \cite{Kalberla2007}. 

\subsection{2MASS} 
\subsubsection{The thin disk}

\citet[][A00]{Alard00} was the first to report on evidence for the flaring of
stars but his contribution remains unpublished. He analyzed three
stripes of 2MASS data oriented perpendicular to the Galactic plane with
$ |b| < 50\degr$ towards the longitudes $l=66\degr, l=180\degr, $ and $
l=240\degr$.  $5.6\, 10^6$ stars in 169 lines of sight were analyzed. 
We plot the results in Fig. \ref{fig1}.

\subsubsection{The old stellar population}

\citet[][LC02]{Lopez02} selected two well-defined
samples for the determination of stellar flaring, red clump giants and
old disk stars, limited in apparent K band magnitude to 14.0. Young
stars and spiral arms were avoided. 
2MASS data were used within distances $ 4 < R < 15 $ kpc.
The basic assumption was that their 
sample is dominated by the old disk population. A relaxed distribution
is expected with well defined scale heights.

Both samples were found to give consistent results. \citet{Lopez02} derive
the best fit flaring for $h_z(R_\odot) = 0.285$ kpc
\begin{equation}
h_z(R) = h_z(R_\odot) exp( (R-R_\odot)/ (12 -0.6 R) ) 
\end{equation}
for $ R < 15$ kpc and conclude that their result agrees well with
\cite{Alard00} for $ R < 12 $ kpc. We plot this fit in Fig. \ref{fig2}.

\subsubsection{Red clump and red clump giant stars}

\citet[][M06]{Momany2006} used 2MASS red clump and red clump giant
stars, selected at fixed heliocentric distances $D = 2.8, 7.5$ and 16.6
kpc. These objects are good standard candles for estimating distances
and were chosen under the assumption that they may suffer least external
contamination by nearby dwarfs. The authors find indications for a
stellar flaring but did not attempt to parametrize it. We therefore show
their data in Figs. \ref{fig2} and \ref{fig3}. While the $D = 2.8 $ kpc
sample appears to be more representative for the thin disk
($h_z(R_\odot) \sim 0.35 $ kpc), the $D = 7.5$ and 16.6 kpc sample shows
a flaring with an intermediate local scale height of $h_z(R_\odot) \sim
0.65 $ kpc, more characteristic for the thick disk.

\subsection{SDSS}

\citet[][H11]{Hammersley2011} used stars of type F8V to G5V from
Sloan-Digital Sky Survey (SDSS) data release DR7
\citep{Abazajian2009}. Five fields for $150\degr < l < 223\degr $ and
$11\degr < b < 31\degr $ were chosen. They modeled the thin disk
assuming an admixture from the thick disk (density 9\% of that of the thin
disk) with a common flare for both components according to
\begin{equation}
h_{z,thin/thick}(R)= 
\left\{  \begin{array}{ll}
        h_{z,thin/thick}(R_\odot) ,& \mbox{ $R\le 16$ kpc} \\
        h_{z,thin/thick} (R_\odot)\exp\left(\frac{R-16}{h_{rf}}\right),
	& \mbox{ $R>16$ kpc.}
\end{array}
\right.
\end{equation}
For $R > 16$ kpc they obtained a flare with a radial scale length
$h_{rf} = 4.5 \pm 1.5 $ kpc. Local thin and thick disk scale heights
were fixed to $h_{z,thin}(R_\odot) = 0.186 $ kpc and
$h_{z,thick}(R_\odot) = 0.631 $ kpc. The corresponding thin 
disk flaring curve is plotted in Fig. \ref{fig2}.

\subsection{SDSS-SEGUE}

\citet[][LC14]{Lopez14} used stars of type F8V-G5V
from the SDSS - Sloan Extension for Galactic
Understanding and Exploration (SDSS-SEGUE) constrained to $ R < 30$ kpc
and $|z| < 15$ kpc. To avoid strong extinction within the Galactic
plane, they used only stars within $8\degr < |b| < 22\degr$. The radial scale
length and the local scale heights for thin and thick disk were fitted
in regions with $ |z| < 3$ kpc, $ R < 15 $ kpc. For a second order flaring
fit all regions with $ 1.5 < |z| < 3.5 $ kpc, $ 7.5 < R < 30 $ kpc were
used. They obtain
\begin{equation}
h_{\rm z, thin}(R) = h_{\rm z, thin}(R_\odot)\left( 1 - 0.037(R-R_\odot
  ) + 0.052(R-R_\odot)^2 \right)
,\end{equation}\[
h_{\rm z, thick}(R) = \left\{ \begin{array}{lcl}
      h_{\rm z, thick}(R_{\rm ft})&  R<R_{\rm ft}\\
      h_{\rm z, thick}(R_{\rm ft})\left( 1 + 0.021(R-R_{\rm ft}) +
        0.006(R-R_{\rm ft})^2 \right) & R \geq R_{\rm ft}\\
   \end{array}
   \right.
.\]
with local scale heights $h_{\rm z, thin}(R_\odot) = 0.24 $
kpc and $h_{\rm z, thick}(R_{\rm ft}) = 0.71 $ kpc for $R_{\rm ft} = 6.9$
kpc. We plot the results in Fig. \ref{fig2} and \ref{fig3}.

\subsection{Pulsars}

\citet[][Y04]{Yusifov2004} analyzed 1412 Pulsars
from the ATNF database \citep{Manchester2005}.  Distances to the pulsars
are estimated from the observed dispersion measure and a model of
the free electron distribution in the Milky Way \citep{Cordes2002}. 
It is currently assumed that such distances are accurate to 30\%.

For $ 5 < R < 18 $ kpc the flare was fitted by 
\begin{equation}
h_z(R) = h_z(R_\odot) exp( (R - R_\odot) / 14. ) 
\end{equation}
with $ h_z(R_\odot) = 0.58 $ kpc, also plotted in  Fig. \ref{fig3}.
  
\section{Comparing flaring data}

In Fig. \ref{fig1} we plot exponential scale heights $h_z$ derived by
\citet[][A00]{Alard00} in comparison to the LAB {H\,{\sc i}} flaring of
\cite{Kalberla2007}. To allow a direct comparison we scale the gaseous
flaring curve to the scale height of the thin stellar disk. A similar
ansatz was implicitly used by \cite{Feast2014} to prepare their
Fig.\,1. This approach (red line in Fig. \ref{fig1}, plotted only for $R
> 12$ kpc) is model independent.

The flaring derived from the best fit mass model by \cite{Kalberla2007}
is marked by the green dashed line. The agreement of the observed
{H\,{\sc i}} scale heights with this mass model is good for $ R < 18$
kpc. At larger distances the {H\,{\sc i}} data show a considerable
scatter with systematic deviations between the northern and southern
hemisphere, while the model is axisymmetric
\citep[][Fig. 4]{Kalberla2008}. For $R < 18 $ kpc the agreement of the
stellar flaring from \citet{Alard00} with the scaled {H\,{\sc i}} data
is excellent, suggesting a common origin for both. Even the dip at $R
\sim 18$ kpc, attributed to the presence of a massive ring in the mass
model of \cite{Kalberla2007}, is consistently visible in both data
sets. The dark matter disk and ring model was constructed to model
details of the {H\,{\sc i}} gas distribution in the Milky Way, in
particular the observed flaring.

The $R , |z|$ positions of the Cepheids \citep[][F14]{Feast2014} were
included in Fig. \ref{fig1} for comparison. They are located
at radial distances 22 to 30 kpc beyond the Galactic center region.
Obscuration may be considerable, which reduces the observability of
Cepheids at lower $|z|$. It needs to be taken into
account that the 2MASS data points represent the scale heights derived
for an ensemble of objects, while the Cepheid data are merely from five
individual objects.  \citet{Alard00} analyzed in total $5.6\, 10^6$
stars, this explains that his data points show a smooth flaring
distribution, essentially without any scatter. 

The \citet{Alard00} sample is by far the largest sample available to us
while the \citet{Feast2014} Cepheid sample is rather restricted but has
well defined distances. Next we discuss the other observations. 

\subsection{A common flaring model for stars and gas?}
Because the best fit mass model derived from the gaseous flaring by
\citet{Kalberla2007,Kalberla2008} is not generally accepted we focus our
discussion in the following on a simple exponential flaring as proposed
by \citet[][Sect. 3.1.4]{KK2009}.
\begin{equation}
h_z(R) = h_z(R_\odot) exp( (R - R_\odot) / 9.22 ) 
\label{EQflare}
\end{equation}
with $R_\odot = 8.0 $ kpc fits the global {H\,{\sc i}} flaring for $5 <
R < 33 $ kpc well. This relation is based on the finding that the
{H\,{\sc i}} distribution in the Milky Way shows an exponential decline
with radius $R$ for both, surface density $\Sigma (R)$ and mid-plane
volume density $n_0(R)$. For such a double exponential distribution,
$h_z(R) \propto \Sigma(R) / n_0(R)$ needs also to be exponential.

If stars flare in a similar way as the {H\,{\sc i}} gas, Equation
\ref{EQflare} should apply to all stellar populations, regardless of the
local scale height $h_z(R_\odot)$. We intend to test this hypothesis.
In the following we distinguish between observations from the
thin and thick stellar disk. According to \citet{Dehnen98} the thin disk
has a scale height of $h_z(R_\odot) = 0.18$ kpc, the thick disk of
$h_z(R_\odot) = 1.0$ kpc. For the further discussion we have chosen to
separate thin and thick disk data according to the gap in scale heights
of our sample at $0.35 \la h_z(R_\odot) \la 0.65$ kpc.

In Fig. \ref{fig2} we plot the exponential approximation for $
h_z(R_\odot) = 0.1, 0.2, 0.3,$ and 0.4 kpc according to Equation
\ref{EQflare} to cover a range that is characteristic for the thin
stellar disk. We compare this with flaring fits according to
\citet[][H11]{Hammersley2011} for $h_z(R_\odot) = 0.186$ kpc,
\citet[][LC14]{Lopez14} for $h_z(R_\odot) = 0.240$ kpc, and
\citet[][LC02]{Lopez02} for $h_z(R_\odot) = 0.285$ kpc. There is a
common trend that stellar flaring at large galactocentric distance $R$
is much stronger than expected from Equation \ref{EQflare}. In
Fig. \ref{fig2} we add data from \citet[][M06]{Momany2006} for
$h_z(R_\odot) = 0.35$ kpc which cover a restricted range only but are
within the uncertainties consistent with the {H\,{\sc i}} flaring.

In Fig. \ref{fig3} we indicate a set of exponential flaring curves for
scale heights of $ h_z(R_\odot) = 0.2, 0.4, 0.6,$ and 0.8 kpc (black
lines). Shown are also observational fits from
\citet[][Y04]{Yusifov2004} (blue dotted), and \citet[][LC14]{Lopez14}
(red).  Data points are from \citet[][M06]{Momany2006}. We find a
general trend that is in opposite to Fig. \ref{fig2}. The observed
flaring at large radial distances falls below the expectations.

From our working hypothesis that stars share the gaseous flaring which
was supported by \citet{Alard00} and \citet{Feast2014} we come to the
striking result that stellar flaring appears to depend on the scale height
$h_z(R_\odot)$ of the objects. Thin disk populations show a very strong
flaring for large distances $R$ while thick disk stars are in this range
below the expectations. 

The velocity dispersion of the {H\,{\sc i} gas is on average
  constant. The exponential flaring curves according to Equation
  \ref{EQflare} apply therefore to the isothermal case $\sigma_z(R) = $
  const. Taking the results from Figs. \ref{fig2} \& \ref{fig3} at face
  value, it is evident from the barometric equation that the velocity
  dispersions $\sigma_z$ for the thin disk populations from our sample
  must {\it increase} with $R$ while for the thick disk populations a
  {\it decrease} is necessary.

It is beyond the scope of this paper to discuss in detail uncertainties
and possible systematical errors in the determination of stellar
flaring. Figure 5 of \citet{Lopez14} may be characterizing the situation
best; the uncertainties are very significant and there is a cross-over for
the thin and thick disk flaring curves at $R \sim 18$ kpc. It is in
general difficult to disentangle thin and thick disk objects at large
radial distances $R$. It is possible that at large distances the thin
disk gets increasingly contaminated by thick disk stars, and opposite
the observed thick disk by thin disk stars. Mixing up the contributions could
lead to biases but it would be a surprise that most of the
investigations that went into Figs. \ref{fig2} and \ref{fig3} are
affected by similar problems.

\section{Discussion}

The question, whether stellar flaring needs to be included in Galactic
mass models is controversial since decades \citep{Rohlfs1981}. The most
recent mass models still assume that the stellar disk has a constant
scale height independent of galactocentric radius $R$
\citep[e.g.][]{Dehnen98,OM00,OM01,Klypin2002,McMillan2011}. Flaring was
explicitly disregarded in recent empirical modeling of the Milky Way
disk \citep{Sharma2013}. The authors state: {\it ``The choice of the
  radial dependence is motivated by the desire to produce discs in which
  the scale height is independent of radius''}. The same constraint was
applied to recent investigations of the kinematic parameters of the
Milky Way disk using the Radial Velocity (RAVE) and Geneva-Copenhagen
(GCS) stellar surveys \citep{Sharma2014}. Here we like to scrutinize the
constant scale height paradigm.

Considering the gaseous components, a major source of uncertainty stems 
from unknown contributions due to the pressure support from magnetic
fields and cosmic rays \citep{Lockman1991}. \cite{Boulares1990} studied
such contributions for a magnetic field perpendicular to the Galactic
plane. On large scales the fields are however found to be predominantly
parallel to the plane \citep{Beck2013}. Therefore we expect that
magnetic fields are important only for local fluctuations in {H\,{\sc
    i}} scale heights, e.g.  related to supernova
remnants. Observational evidence for a common flaring of stars and gas
disfavor the magnetic field hypothesis by \cite{Lozinskaya1963}
immediately.

In summary, we obtain compelling evidence for a common flaring of gas
and stars in the Milky Way galaxy.  {H\,{\sc i}}, Cepheids, 2MASS, SDSS,
and Pulsar data show all consistently increasing scale heights
proportional to $R$. Flaring at large distances $R$ from the Galactic
center appears to be strong for the thin disk but weaker for the thick
stellar disk. Whether or not this indicates distinct different
kinematical properties concerning the associated velocity dispersions
$\sigma_z(R)$ of thin and thick disk stars is an interesting but open
question.

\acknowledgments

We thank Martin Lopez-Corredoira and Yazan Al Momany for distributing
their data and the anonymous referee for constructive
criticism. P.K. and L.D. acknowledge support from Deutsche
Forschungsgemeinschaft, grant KA1265/5, J.K. from grant KE757/7. UH was
supported by institutional research funding IUT26-2 of the Estonian
Ministry of Education and Research.

\clearpage

\begin{figure}
%% \epsscale{.80}
\plotone{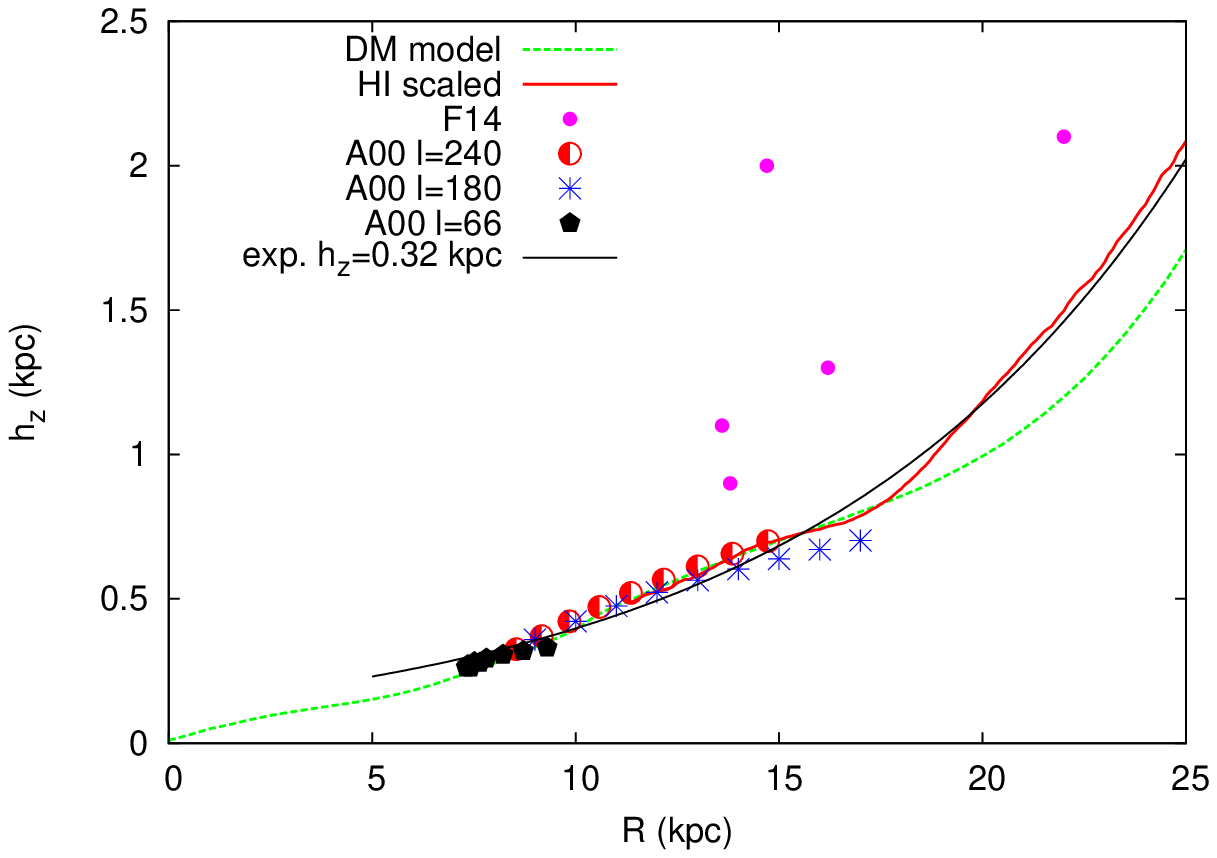}
\caption{Exponential scale heights for the stellar thin disk from
  \cite[][A00]{Alard00} for Galactic longitudes $l = 66\degr$, 180\degr,
  and 240\degr.  The red solid line (plotted only for $R > 12$ kpc) is
  derived from {H\,{\sc i}} data according to \cite{Kalberla2007} after
  matching the gaseous scale heights to that of the stellar component.
  The green dashed line shows flaring curves as expected from {H\,{\sc i}}
  mass model, the black solid line is the exponential
  approximation (\ref{EQflare}). $|z|$ distances of the Cepheids from
  \cite[][F14]{Feast2014} are included for comparison.  }
\label{fig1} 
\end{figure}

\clearpage

\begin{figure}
\plotone{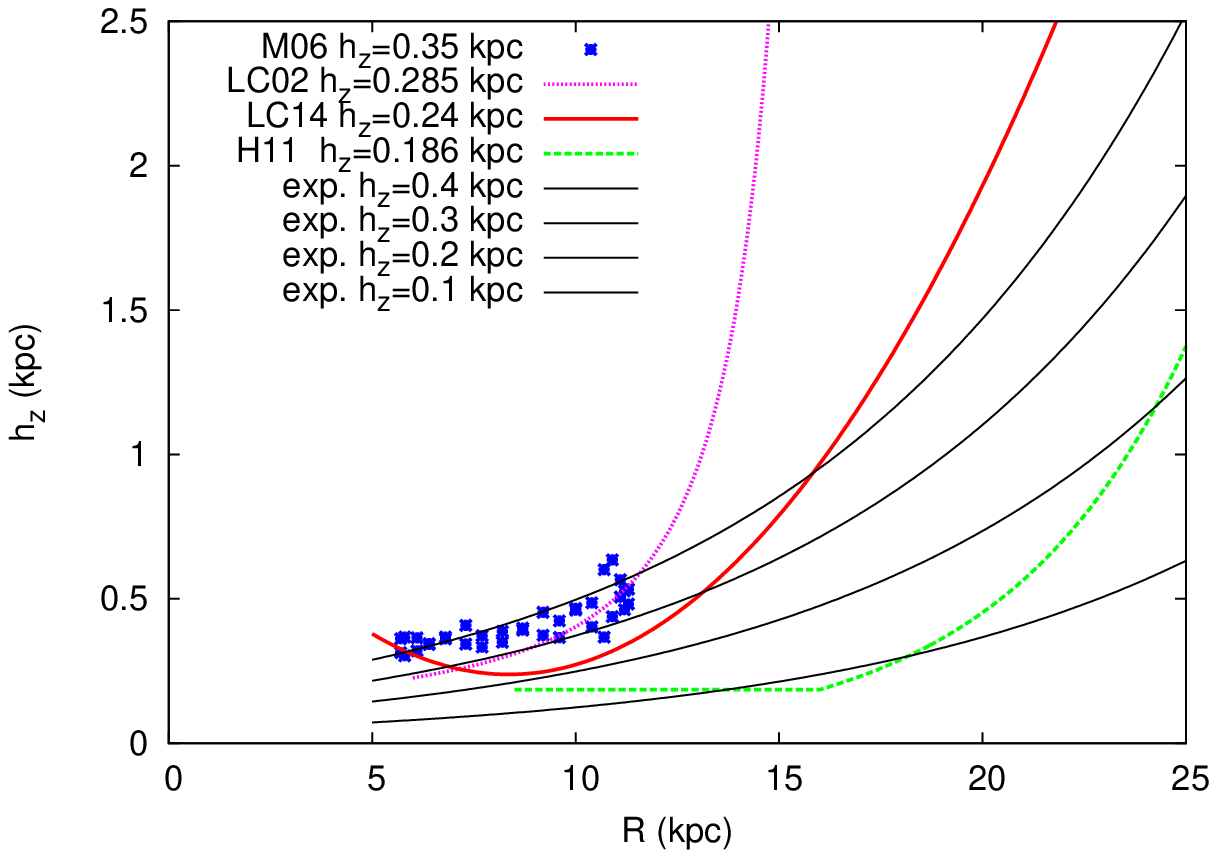}
\caption{ Exponential scale heights $h_z$ for thin disk components. The
  fits are from \citet[][H11]{Hammersley2011} (green dashed),
  \citet[][LC14]{Lopez14} (red), \citet[][LC02]{Lopez02} (magenta dotted),
  Data points in blue are from \citet[][M06]{Momany2006}. The black lines
  represent exponential flaring curves for an isothermal distribution
  with local scale heights of $h_z(R_\odot) = 0.1, 0.2, 0.3,$ and 0.4
  kpc.  }
\label{fig2} 
\end{figure}

\clearpage

\begin{figure}
\plotone{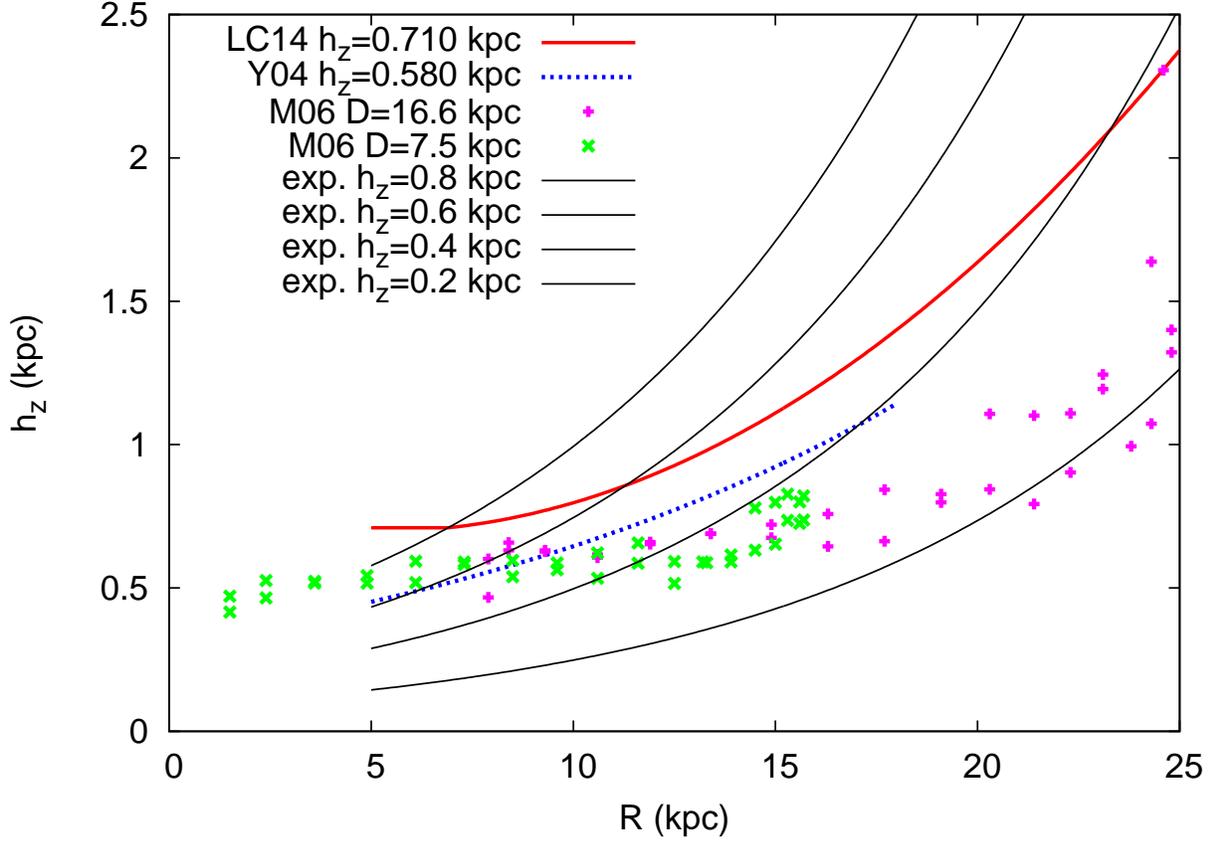}
\caption{Exponential scale heights $h_z$ for thick disk components. The
  fits are from \citet[][LC14]{Lopez14} (red), and
  \citet[][Y04]{Yusifov2004} (blue dotted).  Data points are from
  \citet[][M06]{Momany2006}. The black lines represent exponential flaring
  curves for an isothermal distribution with local scale heights of $h_z =
  0.2, 0.4, 0.6,$ and 0.8 kpc.  }
\label{fig3} 
\end{figure}

\end{document}